\documentclass[letter,twocolumn]{jpsj3}
\usepackage{bm}

\title{Orbital Order, Structural Transition and Superconductivity in
Iron Pnictides}

\author{Yuki Yanagi
\thanks{E-mail address: yanagi@phys.sc.niigata-u.ac.jp}, 
Youichi Yamakawa, Naoko Adachi and Yoshiaki \=Ono}
\inst{Department of Physics, Niigata University, Ikarashi, Niigata
950-2181, Japan \\
}

\abst{
We investigate the 16-band $d$-$p$ model for iron pnictide superconductors 
in the presence of the electron-phonon coupling $g$ 
with the orthorhombic mode which is crucial for reproducing 
the recently observed ultrasonic softening. Within the RPA, 
we obtain the ferro-orbital order below $T_Q$ which induces 
the tetragonal-orthorhombic structural transition at $T_s = T_Q$, 
together with the stripe-type antiferromagnetic order below $T_N$. 
Near the phase transitions, the system shows the $s_{++}$-wave 
superconductivity  due to the orbital fluctuation for 
a large $g$ case with $T_Q>T_N$, while the $s_{\pm}$-wave 
due to the magnetic fluctuation for a small $g$ case with $T_Q \leq T_N$. 
The former case is consistent with the phase diagram of 
doped iron pnictides with $T_s>T_N$. 
}

\kword{iron pnictides, superconductivity, elastic softening, structural transition, orbital order}

\begin{document}
\maketitle

The recently discovered iron pnictide
superconductors\cite{kamihara_1,kamihara} RFe$Pn$O$_{1-x}$F$_x$ 
(R=Rare Earth, $Pn$=As, P) with a transition temperature $T_c$ exceeding
50K have attracted much attention. The parent compounds with $x=0$ 
show the tetragonal-orthorhombic structural transition at $T_s$ and 
the stripe-type antiferromagnetic (AFM) transition at $T_N$. 
The carrier doping $x$  suppresses both of the transition temperatures 
$T_s$ and $T_N$ and induces the superconductivity. In
RFe$Pn$O$_{1-x}$F$_x$, $T_s$ 
is always higher than $T_N$, while in Ba(Fe$_{1-x}$Co$_x$)$_2$As$_2$, 
the simultaneous first-order transition for nondoped case splits 
into two transitions with doping $x$ where $T_s>T_N$\cite{pratt}. 

Theoretically, the $s$-wave pairing with sign change  of the order
parameter 
between the hole  and electron Fermi surfaces (FSs), so called
$s_{\pm}$-wave,
 mediated by the AFM fluctuation was proposed as a
 possible pairing state in the iron pnictides\cite{mazin,kuroki_1, yamakawa_2}. 
The $s_{\pm}$-wave state with a full superconducting gap seems 
to be consistent with most of the experiments\cite{johnston}. 
As for the impurity effects, however, the small $T_c$-suppression 
against nonmagnetic impurities\cite{kawabata_1,sato} is not consistent 
with the  $s_{\pm}$-wave state where $T_c$ is considered to rapidly 
decrease with the nonmagnetic impurities\cite{onari}. Therefore, the
$s$-wave 
state without sign change of the order parameter, so called
$s_{++}$-wave,
 mediated by the orbital fluctuation which is enhanced due to the effects of the 
inter-orbital Coulomb interaction was proposed on the basis of the one-dimensional two-band Hubbard model\cite{sano,okumura} and the two-dimensional 16-band $d$-$p$ model\cite{yamakawa_5}.

\begin{figure}[t]
\begin{center}
\includegraphics[width=8.0cm,angle=-90]{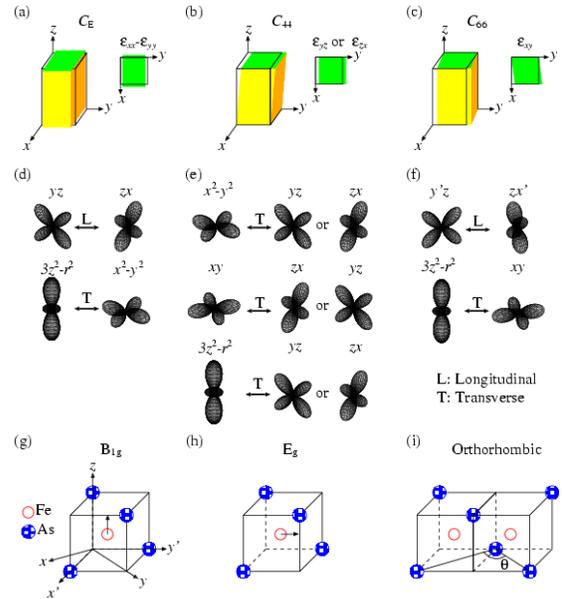}
\caption{(Color online) Schematic figures of the strain fields for
 $C_E$, $C_{44}$ and $C_{66}$ modes (a), (b) and (c),  
 the orbital fluctuations coupled with the corresponding strain fields (d), (e) and (f), 
 and the phonons for $B_{1g}$, $E_g$ and orthorhombic modes 
which enhance the corresponding orbital fluctuations (g), (h) and (i),
 respectively. The $x'$, $y'$ ($x$, $y$) axes are directed
 along the nearest (second nearest) Fe-Fe
bonds.\label{fig_elastic}}
\end{center}
\end{figure}

Remarkably, drastic softenings of the elastic constants have been observed 
in recent ultrasonic 
experiments\cite{mcguire,fernandes,yoshizawa}. As the elastic constant 
$C_\varepsilon$ is given by the second derivative of the total energy
w.r.t. the strain field $\varepsilon$ and includes the contribution such as 
$-g_\eta^2 \chi_\eta$ with the susceptibility $\chi_\eta$ for the
electric operator $\hat{\eta}$ linearly coupled with the strain field 
as $g_\eta \hat{\eta} \varepsilon$, the enhancement of 
$\chi_\eta$ is responsible for the softening of $C_\varepsilon$. 
The detailed ultrasonic measurements\cite{yoshizawa}
 revealed that nondoped  and underdoped Ba(Fe$_{1-x}$Co$_x$)$_2$As$_2$ shows 
drastic softenings of the elastic constants with decreasing $T$ down to $T_s$ 
for $C_E=(C_{11}-C_{12})/2$, $C_{44}$ and $C_{66}$ modes relevant 
to the strain fields $\varepsilon_{xx}-\varepsilon_{yy}$, 
$\varepsilon_{yz}$ ($\varepsilon_{zx}$) and $\varepsilon_{xy}$ 
shown in Fig. \ref{fig_elastic} (a), (b) and (c) which are linearly 
coupled with the orbital fluctuations shown in Figs. \ref{fig_elastic} 
(d), (e) and (f), respectively, where $x', y'$ ($x, y$) axes are 
directed along the nearest (second nearest) Fe-Fe bonds. The softening 
of $C_{66}$ is much larger than $C_E$ and $C_{44}$ and exhibits
divergent behavior when approaching a critical temperature $T_Q$ which 
is just below $T_s$. Then, the orbital susceptibilities relevant to 
$C_{66}$ mode, i. e., the longitudinal $d_{y'z}$-$d_{zx'}$ and/or 
transverse $d_{3z^2-r^2}$-$d_{xy}$ shown in Fig.  \ref{fig_elastic}(f), 
are considered to diverge at $T_Q$ below which the orbital order occurs 
and induces orthorhombic distortion via electron-lattice coupling 
resulting in the structural transition at $T_s \sim T_Q$. 
In addition, optimally doped BaFe$_{1.84}$Co$_{0.16}$As$_2$ shows 
a significant softening of $C_{66}$ mode down to $T_c$\cite{fernandes}. 
Therefore, the orbital order and its fluctuations relevant to $C_{66}$ mode 
are considered to play crucial roles in both the structural transition 
and the superconductivity.

The orbital fluctuation is known to be enhanced by the electron-phonon 
interaction in addition to the inter-orbital Coulomb interaction. 
Recently, the effects of the electron-phonon 
interaction with $B_{1g}$ and $E_g$ modes on the orbital 
fluctuation and its induced $s_{++}$-wave superconductivity 
have been investigated on the basis of the 5-band Hubbard 
model\cite{kontani} and the 16-band $d$-$p$ model\cite{yamakawa_6}. 
As shown in Fig. \ref{fig_elastic}, the $B_{1g}$ phonon enhances 
the longitudinal $d_{yz}$-$d_{zx}$ and transverse 
$d_{3z^2-r^2}$-$d_{x^2-y^2}$ orbital fluctuations responsible 
for the softening of $C_E$ mode, while the $E_g$ phonon enhances 
the transverse $d_{x^2-y^2}$-$d_{yz}$, $d_{xy}$-$d_{zx}$ 
and $d_{3z^2-r^2}$-$d_{yz}$ orbital 
fluctuations responsible for the softening of $C_{44}$ mode. 
However, the effect of the orthorhombic mode which enhances 
the longitudinal $d_{y'z}$-$d_{zx'}$ 
and transverse $d_{3z^2-r^2}$-$d_{xy}$ orbital fluctuations 
responsible for the most dominant softening of $C_{66}$ mode was 
not considered there\cite{kontani,yamakawa_6}. 
The present paper is a straight-forward 
extension of our previous work\cite{yamakawa_6} to include 
the orthorhombic mode which enable us to reproduce the ultrasonic 
softening of $C_{66}$ and to obtain the $x$-$T$ phase diagram 
including the tetragonal-orthorhombic structural transition 
and the superconductivity.

Our Hamiltonian of the two-dimensional 16-band $d$-$p$ Holstein model,  
in which $3d$ orbitals ($d_{3z^2-r^2}$, $d_{x^2-y^2}$, 
$d_{xy}$, $d_{yz}$, $d_{zx}$) of two Fe atoms (Fe$^1$=$A$, Fe$^2$=$B$) 
and $4p$ orbitals ($p_{x}$, $p_{y}$, $p_{z}$) of two As atoms are 
explicitly included, is given by\cite{yamakawa_6}
\begin{equation}
H=H_0+H_\mathrm{int}+H_{ph}+H_{el-ph}, \label{eq_H}
\end{equation}
where $H_0$,  $H_\mathrm{int}$, $H_{ph}$ and $H_{el-ph}$ are the
kinetic, Coulomb interaction, phonon and electron-phonon interaction
parts of the Hamiltonian, respectively.
The kinetic part of the Hamiltonian $H_0$ includes the atomic energies and
the transfer integrals which are 
determined so as to fit both the energy and the
weights 
of orbitals for each band obtained from the
tight-binding approximation to
those from the density functional calculation for LaFeAsO and are listed in
ref. 13. In the present model, 
 the doping concentration $x$ 
corresponds to the number of electrons per unit cell $n=24+2x$ and there
are two hole FSs (FS1 and FS2) around the $\Gamma$ point and two electron FSs (FS3 and
FS4) around the $M$ point for $x=0.1$.
The Coulomb interaction part $H_\mathrm{int}$ includes the 
multi-orbital interaction on a Fe site: the intra- and inter-orbital
direct terms $U$ and $U'$, Hund's rule coupling $J$ and the pair
transfer $J'$. For simplicity, we assume the relation $U=U'+2J$ and
$J=J'$ throughout this paper. 
Hereafter, we number the Fe-$3d$ orbitals as follows:
$d_{3z^2-r^2}$(1), $d_{x^2-y^2}$(2), $d_{xy}$(3), $d_{yz}$(4), 
 $d_{zx}$(5).

\begin{figure}[t]
\begin{center}
\includegraphics[width=8.5cm]{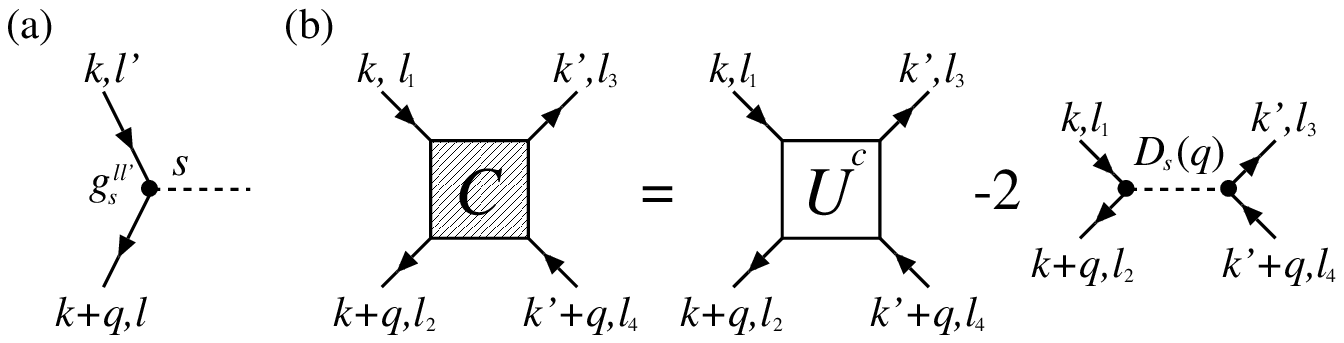}
\caption{Diagrammatic representation of the electron-phonon coupling 
 $\hat{g}_s$ (a) and that of the bare vertex for the charge-orbital 
 susceptibility $\hat{C}$ (b). \label{fig_vertex}} 
\end{center}
\end{figure}

Now we consider the effect of the phonon and the electron-phonon
interaction parts of the Hamiltonian $H_{ph}$ and $H_{el-ph}$ which 
includes the phonon energy $\omega_s$ and the electron-phonon coupling
constant $g^{\ell\ell'}_{s}$ between the orbital $\ell$ and $\ell'$ (see
Fig. \ref{fig_vertex} (a)), 
respectively, where $s$ represents the phonon
mode. 
In the present paper,
we consider the $B_{1g}$,  $E_{g}$ and orthorhombic modes as
shown in Figs. \ref{fig_elastic} (g), (h) and (i).  
We note that the orthorhombic mode is not a normal coordinate but a general coordinate which is given by a linear combination of normal coordinates including both optical and acoustic modes. To avoid the difficulty with many phonon modes, we treat the  orthorhombic mode as a local phonon similar to the $B_{1g}$ and $E_{g}$ modes as a simplest first step in including the orthorhombic mode. 
 As following refs. 17 and 18, we take the electron-phonon coupling
 into account as the atomic energy variance of the Fe-$3d$ electrons.  
The resulting electron-phonon coupling matrix elements are given as
follows: $\sqrt{3}g^{15}_{E_g^1}=g^{25}_{E_g^1}=g^{34}_{E_g^1}
=-\sqrt{3}g^{14}_{E_g^2}=g^{24}_{E_g^2}=-g^{35}_{E_g^2}=g_{E_g}$, 
$g^{44}_{B_{1g}}=-g^{55}_{B_{1g}}=\sqrt{3}/2g^{12}_{B_{1g}}=g_{B_{1g}}$,
$-\sqrt{3}/2 g^{13}_{\theta}=g^{45}_{\theta}=g_{\theta}$, 
$g^{\ell\ell'}_{s}=g^{\ell'\ell}_{s}$ and 0 for otherwise, 
where  $E_{g}^1$ and $E_g^2$
correspond to the oscillation of the Fe atom along the $x$- and $y$-axis, 
respectively, and $\theta$ denotes the orthorhombic mode (see Fig. \ref{fig_elastic}).

Within the RPA\cite{takimoto}, 
the spin susceptibility $\hat{\chi}^s(q)$ and 
the charge-orbital susceptibility $\hat{\chi}^c(q)$ 
are given in the $50\times50$
matrix representation as follows\cite{yamakawa_2,yamakawa_5,yamakawa_6}, 
\begin{eqnarray}
\hat{\chi}^s(q)&=&[\hat{1}-\hat{\chi}^{(0)}(q)\hat{S}]^{-1}\hat{\chi}^{(0)}(q) \label{eq_chis},\\
\hat{\chi}^c(q)&=&[\hat{1}+\hat{\chi}^{(0)}(q)\hat{C}]^{-1}\hat{\chi}^{(0)}(q) \label{eq_chic}
\end{eqnarray}
with the noninteracting susceptibility, $\left[\hat{\chi}^{(0)} (q) \right]^{\alpha,\beta}_{\ell_1\ell_2,\ell_3\ell_4}=
-(T/N)\sum_{k}G^{\beta\alpha}_{\ell_3\ell_1}(k)G^{\alpha\beta}_{\ell_2\ell_4}(k+q)$,
 where  $\alpha$, $\beta$ ($=$$A,B$) represent two
Fe sites, $\ell$ represents Fe 3$d$ orbitals,
$\hat{G}(k)=[(i\varepsilon_n+\mu)\hat{1}-\hat{H}_0(\bm{k})]^{-1}$
is the noninteracting Fe-$3d$ electron Green's function in the
10$\times$10 matrix representation, $\mu$ is the chemical potential, 
$\hat{H}_0(\bm{k})$ is the kinetic part of the Hamiltonian with
 the momentum $\bm{k}$ given in eq. (\ref{eq_H}),
 $k=(\bm{k},i\varepsilon_n)$, $q=(\bm{q},i\nu_m)$ and
$\varepsilon_n=(2n+1)\pi T$ and $\nu_m=2m\pi T$ are the fermionic and
bosonic Matsubara frequencies, respectively. 
It is noted that when  the largest eigenvalue
$\lambda_{\mathrm{spin}}$ ($\lambda_{\mathrm{c-o}}$) of
 $\hat{\chi}^{(0)}(q)\hat{S}$ $(-\hat{\chi}^{(0)}(q)\hat{C})$ reaches unity,
 the magnetic (charge-orbital) instability occurs. In eqs. (\ref{eq_chis}) and
 (\ref{eq_chic}), 
 bare vertices for the spin 
 and charge-orbital 
 susceptibilities $\hat{S}$ and $\hat{C}$  are
 given by\cite{kontani,yamakawa_6}  
$(\hat{S})^{\alpha,\beta}_{\ell_1\ell_2,\ell_3\ell_4}=(\hat{U}^s)^{\alpha,\beta}_{\ell_1\ell_2,\ell_3\ell_4}$, 
$(\hat{C})^{\alpha,\beta}_{\ell_1\ell_2,\ell_3\ell_4}=(\hat{U}^c)^{\alpha,\beta}_{\ell_1\ell_2,\ell_3\ell_4} 
 -2\delta_{\alpha\beta}\sum_{s}g^{\ell_2\ell_1}_{s}g^{\ell_3\ell_4}_{s}D_{s}(i\nu_m)$,  
where $D_{s}(i\nu_m)=2\omega_{s}/(\nu_m^2+\omega_{s}^2)$
is the local phonon Green's function for the
mode $s$ (see Fig. \ref{fig_vertex} (b)).  The bare vertices due to
the Coulomb interaction $\hat{U}^{s(c)}$ are given by,
$(\hat{U}^{s(c)})^{\alpha\alpha}_{\ell\ell,\ell\ell}=U$ $(U)$, 
$(\hat{U}^{s(c)})^{\alpha\alpha}_{\ell\ell',\ell\ell'}=U'$ $(-U'+2J)$, 
$(\hat{U}^{s(c)})^{\alpha\alpha}_{\ell\ell,\ell'\ell'}=J$ $(2U'-J)$ and  
$(\hat{U}^{s(c)})^{\alpha\alpha}_{\ell\ell',\ell'\ell}=J'$ $(J')$, where
 $\ell\ne \ell'$ and the other matrix elements are 0.


The linearized Eliashberg equation is given by 
\begin{eqnarray}
\lambda_{\mathrm{sc}}\Delta^{\alpha\beta}_{\ell\ell'}(k)
\hspace{-2mm}&=&\hspace{-2mm} -\frac{T}{N}\sum_{k'}
\sum_{\ell_1\ell_2\ell_3\ell_4}\sum_{\alpha',\beta'}V^{\alpha,\beta}_{\ell\ell_1,\ell_2\ell'}(k-k')\ \ \ \ \ \ \nonumber\\
&\times& \hspace{-2mm} G^{\alpha' \alpha}_{\ell_3\ell_1}(-k')
\Delta^{\alpha'\beta'}_{\ell_3\ell_4}(k') G^{\beta' \beta}_{\ell_4\ell_2}(k') \label{eq_gap}, \label{gapeq}
\end{eqnarray}
where $\hat{\Delta} (k)$ is the  gap function
and $\hat{V} (q)$ is the
effective pairing interaction for the spin-singlet state.
Within the RPA, $\hat{V} (q)$ 
is given in the  $50\times50$
matrix,
\begin{equation}
\hat{V}(q)=\frac{3}{2}\hat{S}\hat{\chi}^s(q)\hat{S}-\frac{1}{2}\hat{C}\hat{\chi}^c(q)\hat{C}
+\frac{1}{2}\left(\hat{S}+\hat{C}\right)\label{eq_veff_s}.
\end{equation}
The linearized Eliashberg equation (\ref{eq_gap}) is solved to
obtain the gap function $\hat{\Delta}(k)$
with the eigenvalue $\lambda_{\mathrm{sc}}$. At $T=T_c$, the largest eigenvalue $\lambda_{\mathrm{sc}}$ becomes unity.  
We use $32\times32$  
$\bm{k}$ point meshes and 512 Matsubara frequencies  
($-511 \pi T\le \varepsilon_n \le 511\pi T$) in the numerical calculations for
eqs. (\ref{eq_chis})-(\ref{eq_veff_s}). 
For simplicity, we set 
$\omega_{B_{1g}}=\omega_{E_{g}^1}=\omega_{E_{g}^2}=\omega_{\theta}=\omega_0=0.02\mathrm{eV}$
as done in the previous study\cite{yamakawa_6}.  
To reproduce the experimental results that 
the elastic softening is the largest for the 
$C_{66}$ mode\cite{yoshizawa}, 
we assume $g_{B_{1g}}=g_{E_g}=0.85g_{\theta}$ and put $g_{\theta}=g$. 
 Here and hereafter, we
measure the energy in units of eV.

\begin{figure} [t]
\begin{center}
\includegraphics[width=5.5cm]{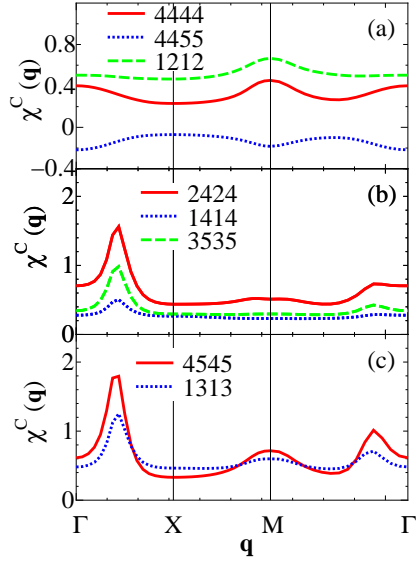}
\caption{(Color online) Several components of the charge-orbital
 susceptibility $\hat{\chi^c}(\bm{q},0)$ for $U'=1.0$, $J=0.2$ and
 $g=0.065$ at $x=0.1$ and $T=0.036$, where 
 we number the Fe-$3d$ orbitals as follows:
$d_{3z^2-r^2}$(1), $d_{x^2-y^2}$(2), $d_{xy}$(3), $d_{yz}$(4), 
 $d_{zx}$(5). 
\label{fig_chic}}
\end{center}
\end{figure}

Fig. \ref{fig_chic} shows several components 
of the static charge-orbital susceptibility $\hat{\chi}^c (\bm{q},0)$
for $U'=1.0$, $J=0.2$ and $g=0.065$ at $x=0.1$ and $T=0.036$. In this case, the dimensionless electron-phonon coupling parameter is given by $\lambda=2g^2\rho_0/\omega_0\sim 2g^2/\omega_0=0.42$ with the density of states at the Fermi level $\rho_0 \sim 1/$eV. 
 We find that, when $T$ decreases, the transverse $d_{yz}$-$d_{zx}$ orbital susceptibility 
 $[\hat{\chi}^c(\bm{q},0)]^{A,A}_{45,45}$, which is equivalent to the longitudinal  $d_{y'z}$-$d_{zx'}$ one, is most enhanced as compared to the other orbital and magnetic susceptibilities (not shown) due to the cooperative effects of the electron-phonon interaction with the orthorhombic mode and the inter-orbital Coulomb interaction $U'$\cite{yamakawa_5,yamakawa_6}. 
We note that the incommensurate peaks around $\bm{q}=(0,0)$ largely 
depend on the electron-phonon coupling strengths $g_s^{\ell \ell'}$ and move to the commensurate peak at $\bm{q}=(0,0)$ for a slightly different parameter set where the resulting pairing state and the phase diagram discussed below is essentially unchanged.



\begin{figure}
\begin{center}
\includegraphics[width=6.0cm,angle=-90]{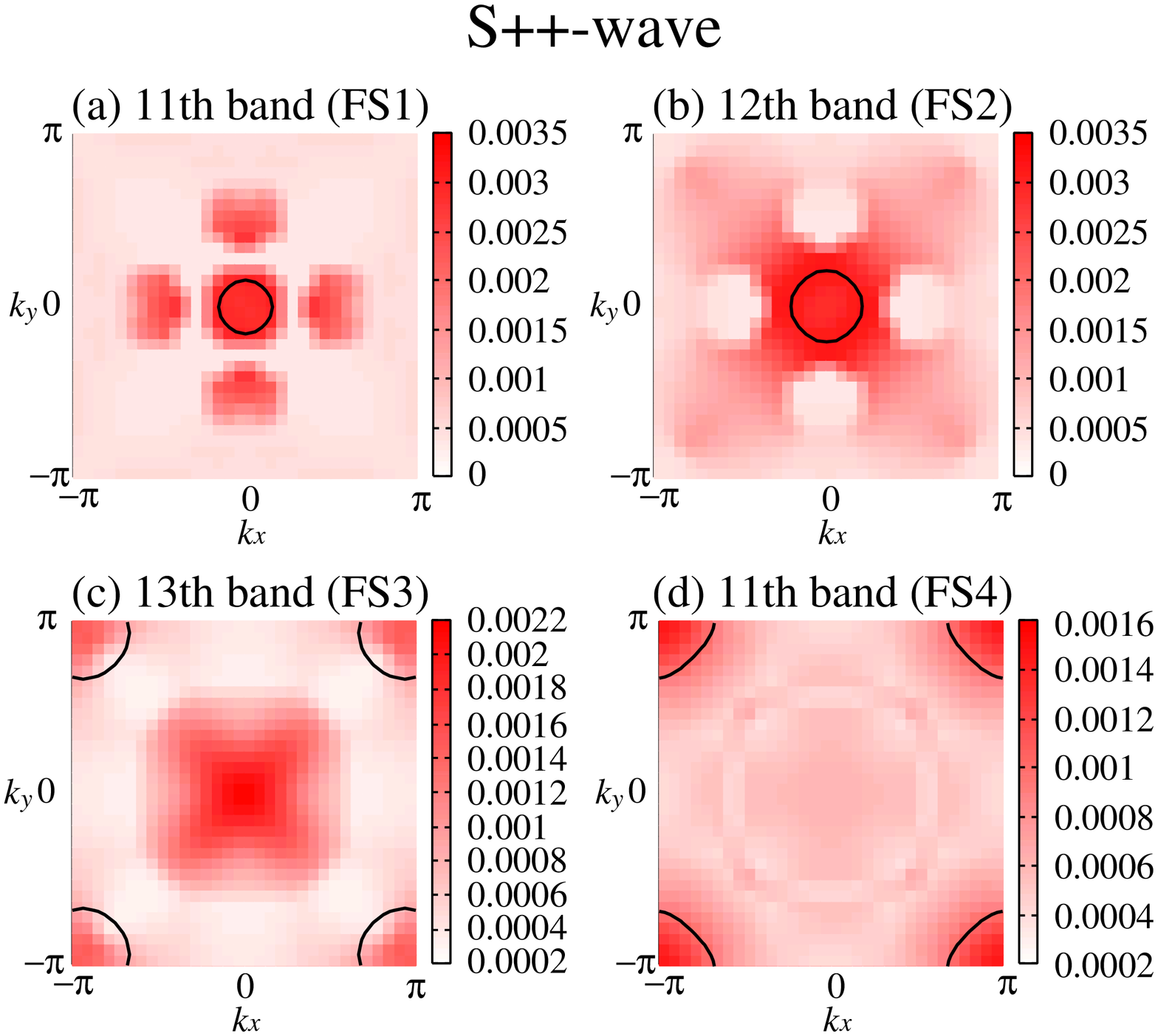}
\includegraphics[width=6.0cm,angle=-90]{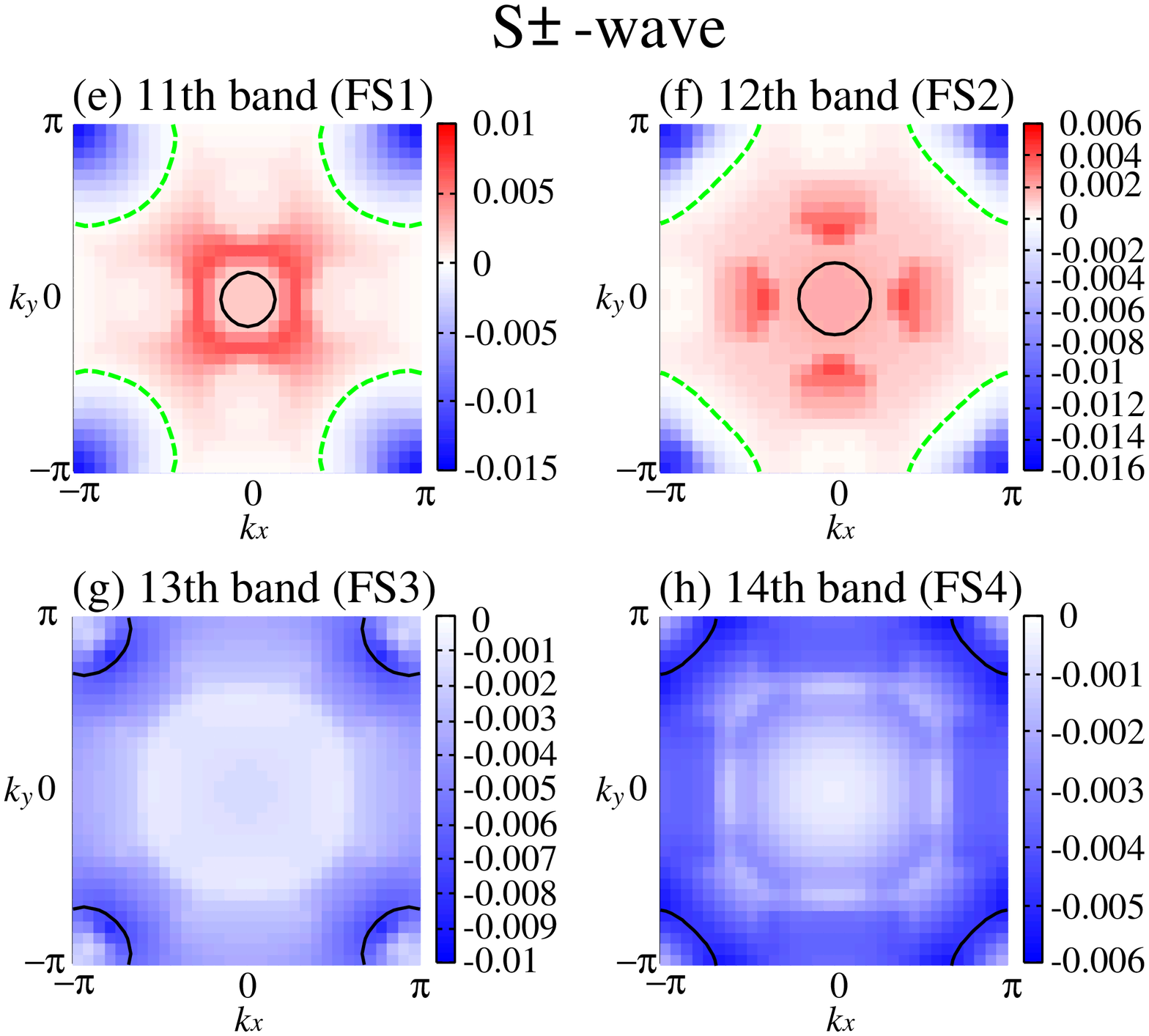}
\caption{(Color online) Several components of the gap function 
$\hat{\Delta}(\bm{k},i\pi T)$ for $U'=1.0$, $J=0.2$ and $g=0.065$
 at $x=0.1$ and $T=0.036$ (a)-(d), and those for  $U'=1.48$, $J=0.2$ and
 $g=0.032$ at $x=0.1$ and $T=0.034$ (e)-(h). 
\label{fig_gap}}
\end{center}
\end{figure}

In Figs. \ref{fig_gap}(a)-(d), we show several components of the gap function with 
the lowest Matsubara frequency $\hat{\Delta}(\bm{k},i\pi T)$ obtained by
solving the linearized Eliashberg equation (\ref{eq_gap}) 
for the same parameters as in Fig. \ref{fig_chic}. In this case, 
the enhanced orbital susceptibility $\hat{\chi}^c(q)$ 
for $\bm{q}\sim (0,0)$ (see Fig. \ref{fig_chic}), i. e., 
the ferro-orbital fluctuation yields the large negative value 
of the effective pairing interactions $\hat{V}(q)$ for 
$\bm{q}\sim (0,0)$ due to the 2nd term of r.h.s. in 
eq. (\ref{eq_veff_s}), resulting in the gap function without sign change, i. e.,  the $s_{++}$-wave state. 
For comparison, we also show the gap function in the case with 
a smaller (larger) value of $g$ ($U'$), $U'=1.44$, $J=0.2$ and
$g=0.032$, in Figs. \ref{fig_gap}(e)-(h). In this case, 
the enhanced magnetic susceptibility $\hat{\chi}^s(q)$ for $\bm{q}\sim
(\pi,\pi)$ (not shown), 
i. e., the stripe-type AFM fluctuation yields the large positive value
of $\hat{V}(q)$ 
for $\bm{q}\sim (\pi,\pi)$ due to the 1st term of r.h.s. in
eq. (\ref{eq_veff_s}), 
resulting in the gap function with sign change, i. e.,  the $s_{\pm}$-wave state. 
We note that, the effects of the ferro-orbital and the AFM fluctuations
on 
the superconductivity do not compete to each other as they are mainly
responsible 
for the different $\bm{q}$ regions in $\hat{V}(q)$, in contrast to the
case with 
the antiferro-orbital and the AFM fluctuations.

\begin{figure}
\begin{center}
\includegraphics[width=6.5cm]{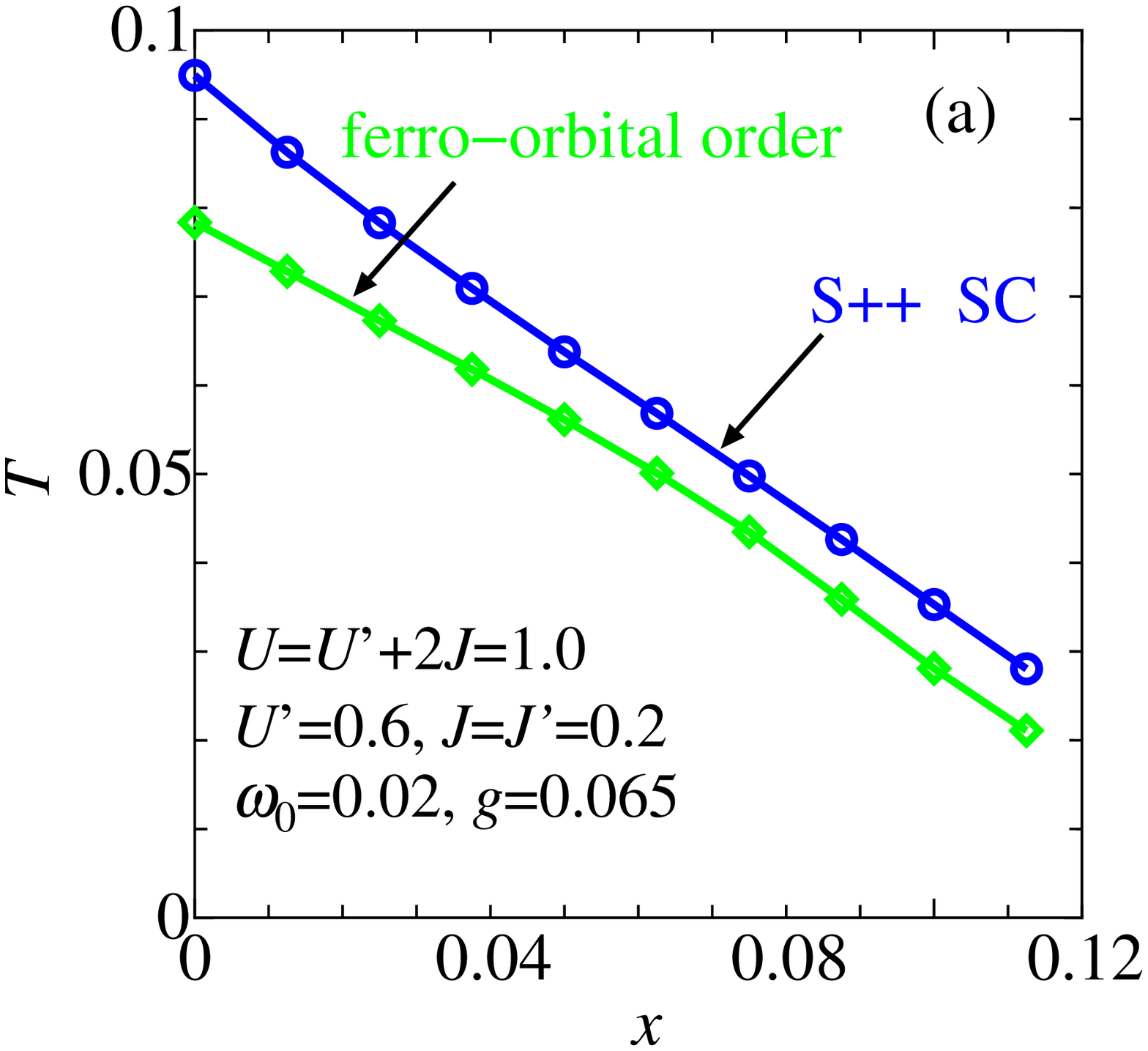}
\includegraphics[width=6.5cm]{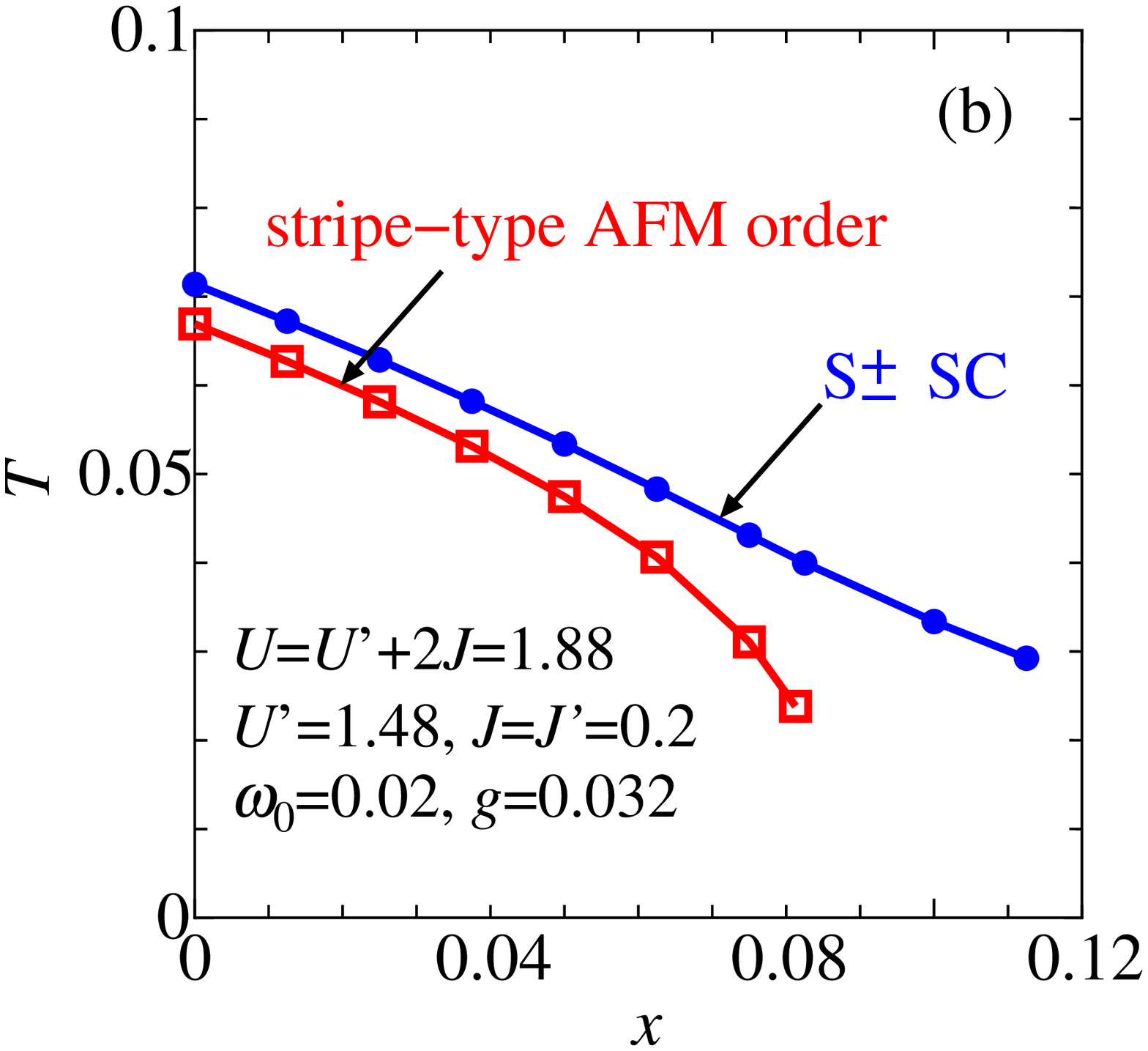}
\caption{(Color online) Phase diagram on the $x$-$T$ plane for $U'=1.0$, $J=0.2$ and $g=0.065$ (a) and that for  $U'=1.48$, $J=0.2$ and $g=0.032$ (b).  
The symbols represent the instabilities for the ferro-orbital order (diamonds), the stripe-type AFM order (squares), the $s_{++}$-wave superconductivity (open circles) and the $s_{\pm}$-wave superconductivity (closed circles), respectively. 
\label{fig_pd_1}}
\end{center}
\end{figure}

Fig. \ref{fig_pd_1}(a) shows the phase diagram on the $x$-$T$ plane in the case with a large $g$, $U'=1.0$, $J=0.2$ and $g=0.065$. When $T$ decreases, the orbital  susceptibilities  $[\hat{\chi}^c(\bm{q},0)]^{A,A}_{45,45}$ and $[\hat{\chi}^c (\bm{q},0)]^{A,A}_{13,13}$ with $\bm{q}\sim (0,0)$ (see Fig. \ref{fig_chic}(c)) diverge at a critical temperature $T_Q$. Below $T_Q$, the ferro-orbital order with different occupations of the $d_{y'z}$ and $d_{zx'}$ orbitals occurs and induces the orthorhombic distortion resulting in the tetragonal-orthorhombic structural transition at $T_s = T_Q$. 
When approaching $T_Q$, the ferro-orbital fluctuation is largely enhanced and mediates the 
$s_{++}$-wave superconductivity (see Figs. \ref{fig_gap}(a)-(d)). 
We also investigate the same model within the Hartree-Fock approximation\cite{adachi} and obtain the phase diagram consistent with RFe$Pn$O$_{1-x}$F$_x$ where $T_s=T_Q$ is always higher than the stripe-type AFM transition at $T_N$ for the case with relatively larger (smaller) orbital (magnetic) fluctuation, and also obtain the phase diagram consistent with Ba(Fe$_{1-x}$Co$_x$)$_2$As$_2$ where the simultaneous first-order transition $T_s=T_Q=T_N$ for $x=0$ splits into two transitions $T_s=T_Q>T_N$ with doping $x$ for a relatively smaller (larger) orbital (magnetic) fluctuation case.  For the both cases, the ferro-orbital fluctuation dominates over the AFM fluctuation above $T_s=T_Q$ for $x>0$ where the $s_{++}$-wave superconductivity is realized.

For comparison, we also show the  $x$-$T$ phase diagram 
in the case with a smaller (larger) value of 
$g$ ($U'$), $U'=1.44$, $J=0.2$ and $g=0.032$, in Fig. \ref{fig_pd_1}(b).  
When $T$ decreases, the magnetic  susceptibility  
with $\bm{q}\sim (\pi,\pi)$ diverges at  $T_N$ below 
which the stripe-type AFM order occurs and induces 
the ferro-orbital order\cite{kubo} together with 
the orthorhombic distortion resulting in 
the tetragonal-orthorhombic structural transition at $T_s = T_N$,
although the RPA result of $T_Q$ is smaller than that of $T_N$.  
When approaching $T_N$, the AFM fluctuation is largely 
enhanced and mediates the $s_{\pm}$-wave 
superconductivity\cite{mazin,kuroki_1, yamakawa_2}. 
In this case, the simultaneous phase transition takes place 
at $T_s=T_N$ even for $x>0$ and is inconsistent with 
the phase diagram of doped iron pnictides with $T_s>T_N$ 
which is reproduced for a large $g$ case mentioned above.

In summary, we have shown that the ferro-orbital fluctuation 
relevant to the ultrasonic softening of $C_{66}$ is enhanced 
by the electron-phonon coupling $g$ with the orthorhombic mode 
and diverges at $T_Q$ below which the ferro-orbital order with 
different occupations of the $d_{y'z}$ and $d_{zx'}$ orbitals 
occurs and induces the orthorhombic distortion resulting in 
the tetragonal-orthorhombic structural transition at $T_s = T_Q$. 
Near the transition, the $s_{++}$-wave superconductivity is 
realized due to the ferro-orbital fluctuation. 
The obtained $x$-$T$ phase diagram is consistent with 
the phase diagram of doped iron pnictides with $T_s>T_N$, 
in contrast to a relatively small $g$ case with $T_s = T_N$ 
where the $s_{\pm}$-wave superconductivity is 
realized due to the antiferromagnetic fluctuation. 
For both cases with $s_{++}$- and $s_{\pm}$-wave superconductivities, the 
RPA result of $T_c$ is always higher than that of $T_Q$ and/or $T_N$,
where the orbital and/or the magnetic fluctuations diverge. 
With including the effects of the self-energy correction and the vertex correction
neglected in the RPA, it is expected that 
the ferro-orbital order and/or the antiferromagnetic orders are realized 
for relatively small $x$, while the
superconductivity is realized for relatively large $x$. The explicit
caluculation including such effects is a future problem. 

The obtained $s_{++}$-wave superconductivity seems to 
be consistent with experimental results of iron pnictides 
including the impurity effects. The enhanced ferro-orbital 
fluctuation above $T_Q$ might be observed by experiments 
with a kind of external field inducing the anisotropy of 
$x'$, $y'$ axes, similar to the case with the ferromagnetic 
fluctuation above the Curie temperature observed by experiments 
with the external magnetic field. In fact, a resistivity anisotropy 
for $T>T_s$ is induced by uniaxial stress\cite{chu}. 
In the present paper, we treated the orthorhombic mode as 
a optical phonon, as a simplest first step. More realistic model 
including acoustic phonons together with a suitable parameter set of 
the electron-phonon coupling strengths $g_s^{ll'}$ will be discussed in a subsequent paper.

\begin{acknowledgments}
The authors thank H. Fukuyama and M. Sato for critical reading of the manuscript and many valuable comments and M. Yoshizawa and H. Kontani for fruitful discussions.   This work was partially supported by the 
Grant-in-Aid for Scientific Research from the Ministry of Education,
Culture, Sports, Science and Technology 
and also by the Grant-in-Aid for JSPS Fellows. 
\end{acknowledgments}


\begin{thebibliography}{10}

\bibitem{kamihara_1}
Y.~Kamihara, H.~Hiramatsu, M.~Hirano, H.~Y. R.~Kawamura, T.~Kamiya, and
  H.~Hosono: J. Am. Chem. Soc. {\bfseries 128} (2006) 10012.

\bibitem{kamihara}
Y.~Kamihara, T.~Watanabe, M.~Hirano, and H.~Hosono: J. Am. Chem. Soc.
  {\bfseries 130} (2008) 3296.

\bibitem{pratt}
D.~K. Pratt, W.~Tian, A.~Kreyssig, J.~L. Zarestky, S.~Nandi, N.~Ni, S.~L.
  Bud{'}ko, P.~C. Canfield, A.~I. Goldman, and R.~J. McQueeney: Phys. Rev.
  Lett. {\bfseries 103} (2009) 087001.

\bibitem{mazin}
I.~I. Mazin, D.~J. Singh, M.~D. Johannes, and M.~H. Du: Phys. Rev. Lett.
  {\bfseries 101} (2008) 057003.

\bibitem{kuroki_1}
K.~Kuroki, S.~Onari, R.~Arita, H.~Usui, Y.~Tanaka, H.~Kontani, and H.~Aoki:
  Phys. Rev. Lett. {\bfseries 101} (2008) 087004.

\bibitem{yamakawa_2}
Y.~Yanagi, Y.~Yamakawa, and Y.~\=Ono: J. Phys. Soc. Jpn. {\bfseries 77} (2008)
  123701.

\bibitem{johnston}
D.~C. Johnston:
 arXiv:1005.4392.

\bibitem{kawabata_1}
A.~Kawabata, S.~C. Lee, T.~Moyoshi, Y.~Kobayashi, and M.~Sato: J. Phys. Soc.
  Jpn. {\bfseries 77} (2008) 103704.

\bibitem{sato}
M.~Sato, Y.~Kobayashi, S.~Lee, H.Takahashi, E.~Satomi, and Y.~Miura: J. Phys.
  Soc. Jpn. {\bfseries 79} (2010) 014710.

\bibitem{onari}
S.~Onari and H.~Kontani: Phys. Rev. Lett. {\bfseries 103} (2009) 177001.

\bibitem{sano}
K.~Sano and Y.~\=Ono: J. Phys. Soc. Jpn. {\bfseries 78} (2009) 124706.

\bibitem{okumura}
M.~Okumura, N.~Nakai, H.~Nakamura, N.~Hayashi, S.~Yamada, and M.~Machida:
  Physica C {\bfseries 469} (2009) 932.

\bibitem{yamakawa_5}
Y.~Yanagi, Y.~Yamakawa, and Y.~\=Ono: Phys. Rev. B {\bfseries 81} (2010)
  054518.

\bibitem{mcguire}
M.~A. McGuire, A.~D. Christianson, A.~S. Sefat, B.~C. Sales, M.~D. Lumsden,
  R.~Jin, E.~A. Payzant, D.~Mandrus, Y.~Luan, V.~Keppens, V.~Varadarajan, J.~W.
  Brill, R.~P. Hermann, M.~T. Sougrati, F.~Grandjean, and G.~J. Long: Phys.
  Rev. B {\bfseries 78} (2008) 094517.

\bibitem{fernandes}
R.~M. Fernandes, L.~H. VanBebber, S.~Bhattacharya, P.~Chandra, V.~Keppens,
  D.~Mandrus, M.~A. McGuire, B.~C. Sales, A.~S. Sefat, and J.~Schmalian:
 arXiv:0911.3084.

\bibitem{yoshizawa}
M.~Yoshizawa, R.~Kamiya, R.~Onodera, and Y.~Nakanishi:
 arXiv: 1008.1479.

\bibitem{kontani}
H.~Kontani and S.~Onari: Phys. Rev. Lett. {\bfseries 104} (2010) 157001.

\bibitem{yamakawa_6}
Y.~Yanagi, Y.~Yamakawa, and Y.~\=Ono: Phys. Rev. B {\bfseries 82} (2010)
  064518.

\bibitem{takimoto}
T.~Takimoto, T.~Hotta, and K.~Ueda: Phys. Rev. B {\bfseries 69} (2004) 104504.

\bibitem{adachi}
N.~Adachi, Y.~Yamakawa, Y.~Yanagi, and Y.~\=Ono:
 presented at Autumn Meet. Physical Society of Japan (Sep. 2010), 25aWG-3.

\bibitem{kubo}
K.~Kubo, and P.~Thalmeier: J. Phys. Soc. Jpn. {\bfseries 78} (2009) 083704.

\bibitem{chu}
J-H Chu, J. G. Analytis, K. D. Greve, P. L. McMahon, Z. Islam, Y. Yamamoto, and I. R. Fisher: Science {\bfseries 329} (2010) 824. 

\end{thebibliography}
\end{document}